\documentclass[namedreferences]{SolarPhysics}
\usepackage{epsfig}
\usepackage[optionalrh]{spr-sola-addons} 
\usepackage{url}
\newcommand{\be}{\begin{equation}}
\newcommand{\ee}{\end{equation}}
\newcommand{\beq}{\begin{eqnarray}}
\newcommand{\eeq}{\end{eqnarray}}
\newcommand{\ietal}{{\it et al.}}
\newcommand{\eg}{{\it e.g.}}
\newcommand{\ie}{{\it i.e.}}

\newcommand{\aeta}[3]{  #1, {\it Astron. Astrophys.} {\bf  #2}, #3.}

\newcommand{\aspj}[3]{  #1, {\it Astrophys. J.} {\bf  #2}, #3.}

\newcommand{\jogr}[3]{  #1, {\it J. Geophys. Res.} {\bf  #2}, #3.}

\newcommand{\sph}[3]{   #1, {\it Solar Phys.} {\bf  #2}, #3.}

\newcommand{\pasj}[3]{#1, {\it Publ. Astron. Soc. Japan} {\bf  #2}, #3.}

\begin{document}
\begin{article}
\begin{opening}

\title{Large-Amplitude Oscillation of an Erupting 
Filament as Seen in EUV, H$\alpha$ and Microwave Observations}

\author{H. \surname{Isobe}$^{1}$
        D.   \surname{Tripathi}$^{2}$
        A. \surname{Asai}$^{3}$
        R. \surname{Jain}$^{4}$}

\runningauthor{H. Isobe et al.}
\runningtitle{}

\institute{$^{1}$ Department of Earth and Planetary Science, University 
             of Tokyo, Hongo, Bunkyo-ku, Tokyo 113-0033, Japan
                  \email{isobe@eps.s.u-tokyo.ac.jp}\\
           $^{2}$ Department of Applied Mathematics and Theoretical 
                  Physics, University of Cambridge, 
                  Wilberforce Road, Cambirdge CB3 0WA, UK\\
           $^{3}$ Nobeyama Radio Observatory, National Astronomical 
                  Observatory of Japan, Minamimaki, Minamisaku, 
                  Nagano 384-1305, Japan
                  \\
           $^{4}$ Department of Applied Mathematics, University 
                  of Sheffield, Hicks building, Hounsfield Road, 
                  Sheffield S3 7RH, UK
             }

\date{Received ; accepted }

\begin{abstract}
We present multiwavelength observations of a large-amplitude 
oscillation of a polar crown filament on 15 October 2002, 
which has been reported by Isobe and Tripathi 
({\it Astron. Astrophys.} {\bf 449}, L17, 2006). 
The oscillation occurred during the slow rise ($\approx$ 1 km s$^{-1}$)
of the filament. It completed three cycles 
before sudden acceleration and eruption.  
The oscillation and following eruption were clearly seen 
in observations recorded by the Extreme-Ultraviolet Imaging 
Telescope (EIT) onboard the Solar and Heliospheric Observatory (SOHO). 
The oscillation was seen only in a part of the filament, 
and it appears to be a standing oscillation rather than 
a propagating wave.  
The amplitudes of velocity and spatial displacement of the 
oscillation in the plane of the sky were about 5 km s$^{-1}$ 
and 15,000 km, respectively. The period of oscillation was about two hours 
and did not change significantly during the oscillation. 
The oscillation was also observed in H$\alpha$ by the 
Flare Monitoring Telescope at Hida Observatory. We determine the 
three-dimensional motion of the oscillation from the 
H$\alpha$ wing images. The maximum line-of-sight velocity 
was estimated to be a few tens of km s$^{-1}$, though the uncertainty 
is large owing to the lack of the line-profile information. 
Furthermore, we also identified the spatial displacement of 
the oscillation in 17 GHz microwave images from 
Nobeyama Radio Heliograph (NoRH). 
The filament oscillation seems to be triggered by 
magnetic reconnection between a filament barb and 
nearby emerging magnetic flux as was evident from the 
MDI magnetogram observations. 
No flare was observed to be associated with the onset of the oscillation.
We also discuss possible implications of the oscillation 
as a diagnostic tool for the eruption mechanisms. 
We suggest that in the early phase of eruption a part of the filament lost 
its equilibrium first, while the remaining part was still in an equilibrium and 
oscillated. 
\end{abstract}

\end{opening}

\section{Introduction}
\label{Introduction} 

A variety of oscillatory motions of prominences and filaments have been 
observed and used to infer their structure and physical parameters 
(see the review by Ballester, 2006). 
In particular, global oscillations of filaments initiated 
by flare-associated disturbances (Ramsey and Smith, 1966) 
may provide useful information on the properties of filaments 
as oscillators, and hence on their stability. 

Isobe and Tripathi (2006; hereafter paper I) reported an event 
in which a polar-crown filament exhibited large amplitude oscillation 
in its pre-eruptive slow rise phase. 
This event provides a rare opportunity to use so-called 
coronal seismology or prominence seismology 
in order to study the eruption mechanism of filaments. 
In paper I we reported that the oscillation repeated for about three 
cycles without any significant damping, until the onset of rapid acceleration 
and eruption of the filament. The period and the velocity amplitude 
measured in the plane of sky 
of the oscillation were about 2 hours and 4.2 km s$^{-1}$, respectively. 
It was suggested that the large-amplitude oscillation in the 
filament was evidence for it retaining a non-linear stable 
equilibrium in its pre-eruptive, slow-rise phase. The 
transition from this equilibrium to eruption, \ie, loss of 
equilibrium (LOE) or instability, was triggered by rapid ($\approx$ 
Alfv\'{e}n time scale) mechanism such as magnetic reconnection. 

In this paper we present further detailed analyses of this event using 
multiwavelength observations. 
In addition to the detection of the oscillation in extreme ultraviolet  
and H$\alpha$, we discovered that the oscillation was seen 
also in microwave. 
To the best of our knowledge, this is the first evidence of spatially resolved, 
large-amplitude oscillation of a filament in microwave imaging. 
One of the fundamental issues that remained unknown in paper I was 
the physical cause behind excitation of the oscillation. 
It is known that large amplitude filament oscillations are 
usually excited by flare-associated disturbances 
 (see \eg, Ramsey and Smith,  1966; Eto \ietal,  2002; Jing \ietal, 2003, 2006; 
Okamoto \ietal, 2004; Vr\v{s}nak \ietal, 2007). 
However, no flare associated with the onset of oscillation in this event was observed. 
In this paper we examine the magnetic-field data to study the mechanism 
of the excitation of the oscillation.  

The paper is organized as follows: In Section 2 
we briefly summarize the observational data. In Section 3 
the oscillation seen in various wavelengths is presented. 
We discuss the possible excitation mechanisms responsible for the 
filament oscillation and their implication to eruption mechanisms in Section 4. 
Conclusions are summarized in Section 5.

\section{Observations}
\label{observations}
The oscillation and subsequent eruption of a filament was  
observed on 15 October 2002. The filament was a large polar 
crown in the southern polar region. 
The oscillation was most clearly seen in 195 \AA \, images 
taken by the Extreme-Ultraviolet Imaging Telescope 
(EIT; Delaboudini\`{e}re \ietal, 1995) onboard the 
{\it Solar and Heliospheric Observatory} (SOHO).
The cadence and spatial resolution of EIT are  
12 minutes and 2.6 arcsec, respectively. 
The 195~{\AA} passband of the EIT is dominated by an 
Fe \, {\sc xii}  
line formed at 1\,--\,2 MK, but also 
contains an Fe \,{\sc xxii} line at 192~{\AA} formed at around 
20 \,MK, which is usually much weaker in the quiet Sun 
region but highly significant in flaring active region (Tripathi \ietal, 2006). 
The filament (cold plasma in the corona) is seen as a dark feature due to 
absorption in the H\,{\sc i} Lyman continuum (Kucera, Andretta and Poland, 1998; 
Engvold \ietal, 2001). 
We also used the line-of-sight magnetograms recorded by the Michelson Doppler Imager 
(MDI: Scherrer \ietal, 1995) onboard SOHO to study the photospheric magnetic-field configuration. 

The filament oscillation was also observed by the Flare 
Monitoring Telescope (FMT) at Hida Observatory, Kyoto University.
FMT observes the full disk of the Sun in five channels including 
H$\alpha$ center and wings at $\pm 0.8$ \AA \, with the time 
cadence of two seconds and a pixel sampling of 4.25 arcsec 
(Kurokawa \ietal, 1995). The advantage of FMT is its continuous, 
full disk observations in H$\alpha$ wings. It has been used 
for the studies of transient and dynamic events 
such as filament oscillation (Eto \ietal, 2002; Okamoto \ietal, 2004), 
filament eruptions (Morimoto and Kurokawa, 2003b) and surges 
(Liu \ietal, 2005). 

The spatial displacement associated with the oscillation was also identified 
in the microwave observation at 
17 GHz from Nobeyama Radio Heliograph (NoRH; Nakajima \ietal, 1994). 
The advantage of microwave observation is that it is insensitive 
to temperature and velocity of filaments/prominences, \ie, 
the visibility of the filament does not change even when 
heating or motion of the filament occur. Hence the NoRH 
data have been used for the studies of prominence eruptions 
(Gopalswamy \ietal, 2003; Shimojo \ietal, 2006)

All of the images shown in this paper were rotated to the 
reference time of 06:00 UT to compensate for the effect 
of solar rotation. We used the \url{DROT_MAP} procedure in the Solar SoftWare 
to correct the rotation, and neglect the projection effect caused by 
the height of the filament above the photosphere.

\section{Results}
\subsection{Oscillation and eruption as seen by EIT}
\label{subsec:eitosc}

Figure 1 shows the EIT images of the filament during  
the oscillation (left), around the time of the onset of eruption (middle) 
and after the eruption (right). The $x$ and $y$-axes are 
the distances from the disk center. 
Before the eruption, the eastern half ($x < -250$ arcsec) 
of the filament was seen above the limb. 
(Therefore, part of the filament was actually seen as a ``prominence", 
though we use the term of ``filament" to refer to the whole entity.) 
The western half ($x > -250$ arcsec) was seen as a dark filament on the disk. 
The apparent morphology of the filament is different 
between the eastern and western parts, likely because of projection effects. 
It seems that the eastern 
part is a vertical slab seen from the side, and the western  
part has similar structure but is seen from the top. 
After the eruption, brightenings at the eastern 
and western ends of the filament (right panel: 
[$x,y$]=[-650, -500] and [-100, -500]) are evident. 

The oscillatory motion of the filament can be visualized by 
the temporal slices along the slits perpendicular to the filament. 
Figure 2 shows the temporal slices along 
seven different slits shown in the left panel of Figure 1. 
The right-bottom panel shows the time slice for slit three 
but for the microwave image; see Section 3.2.  
The oscillation starts at around 02:30 UT and completes 
three cycles with a period of about two hours. 
During the fourth cycle the filament is suddenly accelerated 
and erupted. 

The oscillation is most prominent at slit three. 
The amplitude of the oscillation is smaller in the slits 
away from slit three in both directions and is almost unrecognizable in slit seven.  
The vertical black lines indicate the times of the 
first peak (04:00 UT) and the second peak (06:00 UT) 
as determined by visual inspection from the time slice of slit three.  
Although there are some ambiguities in determining the peaks, 
the oscillatory motion at the different slit position  
appears to be in the same phase. Therefore, the oscillatory 
motion can be interpreted as a standing oscillation. 
The amplitude of velocity and spatial displacement are 
about 5 km s$^{-1}$ and 15,000 km, respectively. 
Note that, however, these values are measured in the plane of the sky. 

In Paper I it was suggested that there may be an increase in 
the period based on the visual inspection of the time slices, 
although this was not conclusive. To study the 
change of period  accurately, a wavelet analysis was 
performed on the EIT data; no substantial increase in the period 
was found (Pint\'{e}r \ietal, 2007). 
This result has a significant influence on the diagnosis of the trigger mechanism for the 
filament eruption (see Section \ref{dis-erup}).

The remarkable point of this event is that the oscillation 
occurred while the filament is slowly rising toward the eruption. 
The slow rise motion is seen in Figure 2 as a general trend 
toward the southwest with a velocity of about 1 km s$^{-1}$ 
(See also Figure 2 of paper I). Let us examine the effect of the 
correction of rotation on the velocity of the filament motion. 
As previously mentioned, we neglected the height of the filament 
when correcting the solar rotation. The error caused by this simplification 
is $\lesssim H\theta$, where $H$ is the filament height and $\theta$ is 
the rotated angle. During the 12 hours in which we made the time slices, 
the photosphere at this latitude rotates about 5\,--\,6 degrees. If $H=10^5$ km, 
which seems reasonable, the error in terms of the velocity is about 0.25 km s$^{-1}$ 
or less. Thus the error is smaller than the velocities of the slow rise and 
the oscillation.

\subsection{NoRH and FMT observation of oscillation}
Figure 3 shows the H$\alpha$ line center image from FMT (left panel) 
and the 17 GHz image from NoRH (right panel). 
Unfortunately the eruption occurred 
during the night for FMT and NoRH, but the first period of the oscillation 
was within the observation time of these instruments. 
The white line in the microwave image (right panel) shows 
the position of slit three shown in Figure 1 The time slice of 
this slit is shown in Figure 2 (panel h) which reveals a   
clear oscillatory motion. 

Figure 4 shows the time series of H$\alpha$ images at line center 
and blue (-0.8 \AA) and red (+0.8 \AA) wings. The field of 
view is shown by the rectangle in Figure 3. 
The irregular 
time cadence is due to frequent interruptions by clouds. 
During 04:00 -- 05:00 UT the filament 
appears in the blue wing images as a dark feature. The filament 
fades away in the blue wing around 05:00 UT, and then it appears 
in the red wing (05:02 and 05:27 UT). This is the well-known 
feature of oscillating filaments called ``winking filament''
(Hyder, 1966). 

The three-dimensional velocity field can be inferred 
by comparing the H$\alpha$ wing images and the motion  
in the plane of the sky. When the filament appears in the blue (red)  wing, 
which means it is moving toward (away from) the Earth, 
it is moving toward Northeast (Southwest) in the plane of the sky 
(see Figure 2). 
Given that the filament is located near the southern limb 
of the Sun, if the direction of the oscillatory motion is vertical, 
the blue (red) shift should accompany southward (northward) motion. 
Since in this case the sense of oscillation appears to be opposite, 
we conclude that the oscillatory motion is horizontal. 
The contrast of the wing images is largest (\ie, the filament is dark)  
in the central area of the field of view, where the plane-of-the-sky 
velocity is also largest; see the position of slit 3 shown 
in the right panel of Figure 4. This is consistent with the 
interpretation as a standing oscillation. 

The line-of-sight velocity of the filament can be inferred
from the contrast in the wing images. Morimoto and Kurokawa (2003a) 
developed a method to determine the line-of-sight velocity 
from FMT data based on a cloud model. Using their method, 
we estimated the line-of-sight velocity of the 
oscillation to be about 20\,--\,30 km s$^{-1}$ (see Paper I). 
This value depends on the line profile of the filament, 
which is unknown here (whereas in paper I we assumed a typical line profile 
presented in Morimoto and Kurokawa, 2003a). 
Therefore the uncertainty in the line-of-sight velocity is large.

\section{Discussion}
\subsection{What is the trigger of the oscillation?}
\label{efr}
It is known that large-amplitude oscillations of filaments like 
this event are excited for flare-associated disturbances, such as 
Moreton waves (Ramsey and Smith, 1966; Eto \ietal, 2002), 
EIT waves (Okamoto \ietal, 2004), and nearby subflares (Jing \ietal, 2003, 2006). 
There was no observed flare or brightening associated with the present event that 
can account for the excitation of the oscillation. 
However, we identify an emerging flux region that may play a role 
in triggering the oscillation, and possibly in triggering the eruption as well. 

Figure 5 shows EIT images taken just before the onset of the oscillation. 
An MDI magnetogram taken at 03:12 UT is overlaid on 
the EIT image taken at 02:48 UT shown in the left panel. 
The blue and red contours correspond to +25 and -25 gauss, respectively. 
As indicated in the figure, there is a bipole whose opposite polarities are 
connected by a bright EUV loop.
The lower panels of Figure 5 show temporal evolution of the 
magnetogram in the black box shown in the upper left panel. 
The bipole is already visible at 19:15 UT, October 14, and evolves 
with time as see in Figure 5. The separation of the two polarities 
is the characteristic of an emerging flux. 

Also indicated in Figure 5 is a filament 
barb connected to the preexisting positive polarity region 
near the negative polarity of the emerging flux (02:48). 
This barb was seen for hours before the onset of oscillation. 
At 03:12 UT it suddenly disappears from the EIT images (see the middle panel of Figure 5) 
and a very faint jet-like ejection was seen at 03:36 UT toward the southwest. 
This ejection is co-temporal with the 
onset of oscillation and also co-spatial with the part of the 
filament where the amplitude of the oscillation is largest.  
We interpret these observations as follows. 
As the new flux emerged and expanded into the corona,  it interacted 
with the filament barb and magnetic reconnection occurred. 
This reconnection produced the plasma ejection, which blew the filament 
from the side and excited the oscillation. 
We note that since we could not identify the polarity of the footpoint of the barb in 
the magnetogram, we do not know whether the direction of the 
magnetic field was favorable for reconnection or not. 
However, three-dimensional MHD simulations suggest that 
reconnection between an emerging flux and a coronal field can occur 
regardless of their mutual orientation, though the resultant dynamics and 
magnetic field configuration vary 
(Archontis \ietal, 2005; Galsgaard \ietal, 2007). 

There was no brightenings in EIT images associated with the 
ejection, probably because the released energy is too small to 
brighten EUV or the reconnection occurred in the chromosphere. 
It is also possible that the time scale of brightening associated with the 
reconnection was shorter than the EIT cadence, and the brightening 
was not recorded by the EIT.

\subsection{Internal structure of the filament}
So far we considered the filament as a single entity. However, the filament 
has some internal structure, as seen in the EIT images in Figure 5. One can 
also see internal structure of the filament structure in Figure 3,  which may indicate the 
presence of substructures that could be oscillating a little out of phase. 
Indeed, a movie of EIT images shows some motion along the filament. 
This is probably due to the direction of the perturbation that excited the 
oscillation (Jing \ietal, 2006; Vr\v{s}nak \ietal, 2007). As indicated by the long arrow in Figure 5,   
the direction of the ejection that blows the filament was oblique to the filament axis, 
which possibly excited both lateral and longitudinal motions in the filament.

\subsection{What is the implication for eruption mechanisms?}
\label{dis-erup}
The large-amplitude oscillation occurred while the filament is 
slowly rising before eruption. Slow-rise motions before 
the fast acceleration and eruption have been commonly observed 
in various types of solar plasma ejections, such as filament 
eruption (Sterling and Moore, 2003, 2005; Chifor \ietal, 2006), 
X-ray plasmoid ejections (Ohyama and Shibata, 1997), 
and coronal mass ejections (Zhang \ietal, 2001). 
A fast-acceleration phase following the slow-rise phase has also been found 
in numerical simulations (\eg, Magara, Shibata, and Yokoyama, 1997; 
Chen and Shibata, 2000). 
The temporal evolution of the height of the ejecta seen in observations is similar 
to those in simulations where the ejecta already lost their stable equilibrium 
(for comparison of observation  and simulation, see, \eg, 
Magara, Shibata, and Yokoyama, 1997; 
Sterling and Moore, 2005). 
This is suggestive of an initial growth of an instability or LOE as 
an explanation of the slow rise. 
Such slow rise usually lasts several tens of the Alfv\'{e}n time of the system
(\ie, time in which Alfv\'{e}n waves travel across the system), 
typically $\approx 10^3$ seconds for active region events and 
several hours for quiet region filaments. 

However, a slow rising motion of a filament that lasted much longer 
(period $\approx$ days) before its eruption has been reported by Nagashima \ietal (2007).  
They interpreted  the motion as a quasi-static evolution toward the critical height 
at which the filament lost its equilibrium (Forbes and Isenberg, 1991).  
The quasi-static evolution (a series of MHD equilibria) may be driven by 
photospheric motions (Forbes and Priest, 1995), an emerging flux 
(Lin, Forbes, and Isenberg, 2001), 
or continuous small scale reconnection in the lower atmosphere (Nagashima \ietal, 2007). 
This kind of slow rise is physically different from the former type,  
that is, initial growing phase of the LOE or instability. 

The oscillation during the slow-rise phase provides strong evidence 
that the prominence retained its equilibrium during that time, which 
indicates that the slow rise found in our event is of the latter type: 
the quasi-static evolution {\it before}  the LOE or instability sets in. 
However, our event also shows characteristics to those of the former type, 
the initial growing phase {\it after} the LOE or instability sets in. 
The height-time plot of the filament is quite similar to those  
of the former type; it shows a smooth transition from slow rise with constant 
velocity to the fast acceleration (see time slice for slit seven in Figure 3 and Figure 2 
of Paper I).  The duration of slow rise in our event ($>$ 12 hours) seems 
quite long as that of the former type, but it is probably due to the long Alfv\'{e}n time 
($\sim$ tow hours, see Paper I). 

The oscillation continued until the onset of the fast acceleration. 
The amplitude of the oscillation was large; the displacement in the plane of the 
sky ($\approx$ 20000 km) is as large as the width of the filament. Therefore, during 
the oscillation the equilibrium of the filament, or at least its oscillating part, 
was nonlinearly stable to the large perturbation. 
However, if the eruption occurs through the exchange of stability, we expect that the period 
of the oscillation becomes longer and longer as the system changes its equilibrium state 
and approaches the eruption. 
As mentioned in Section \ref{subsec:eitosc}, the wavelet analysis showed 
no evidence for an increase in period (Pint\'{e}r \ietal, 2007). 
Therefore it is unlikely that the slow rise is due to the change of the equilibrium state 
toward the critical point at which the LOE or instability sets in. 
Moreover, the transition from the nonlinearly 
stable equilibrium to eruption occurred within a period of the oscillation or, in other words, 
in the Alfv\'{e}n time scale. 
We note that emerging flux and its reconnection with the filament barb that 
triggered the oscillation may have played a role in triggering the eruption, too
(Feynman and Martin, 1995; Chen and Shibata, 2000; 
Lin, Forbes, and Isenberg, 2001; Tripathi, 2005). 

However, the following question remains:  
Does the oscillation found in our event suggest that the similar slow rises found commonly in 
other events (Ohyama and Shibata, 1997, Zhang \ietal, 2001; Sterling and Moore, 2003, 2005; 
Chifor \ietal, 2006) also indicate quasi-static evolution, in contrast to the previous interpretation 
as the initial phase of LOE or instability? This may be the case, but 
here we propose another scenario. 

We note that the oscillation occurred only in a part of the filament. The size of 
the oscillating part of the filament is about $2\times10^5$ km (approximately 
the size of the box in Figure 3), whereas the size of the complete erupting filament 
is as large as the solar radius. In such a long filament, it is probably possible that 
the filament loses its equilibrium in a ``global" sense, but a part of the 
filament still retains ``local" equilibrium that can support ``local" oscillation. 
Indeed, in many filament eruption events the whole filament does not erupt 
simultaneously, but often a part of the filament erupts first and the remaining 
parts seem to be dragged by the leading part (a process called ``asymmetric eruption" by 
Tripathi, Isobe, and Mason, 2006. See also Isobe, Shibata, and Machida, 2002). 
Chifor \ietal (2006) analyzed X-ray, EUV, and microwave brightenings of 
such an asymmetric eruption and found  (i) that the filament started to rise 
from one end where an X-ray precursor was found, and (ii) there was a weak heating of the 
filament during the slow rise. They concluded that at first a part of the filament 
started to rise owing to the  localized magnetic reconnection below the filament.  
Then the rise of the filament induced successive magnetic reconnection 
along the filament, and eventually the whole filament was destabilized and erupted 
(see  also Chifor \ietal, 2007). This scenario naturally explain how a localized 
perturbation (such as an emerging flux) eventually leads to the destabilization 
of the whole filament. 

We suggest that the observation presented in this paper 
further supports the model of Chifor \ietal, (2006). 
The oscillation in a part of the filament during the slow rise is evidence 
that that part of the filament still retained an equilibrium. However, the other 
nonoscillating part had already lost the equilibrium and started to rise, dragging 
the oscillating part.  Thus the oscillating part was also forced rise, and 
eventually erupted along with the other parts of the prominence.

\section{Conclusions}
In this paper we presented the observations of a large amplitude 
oscillation of the polar crown filament in its pre-eruption, 
slowly rising phase. The oscillation was seen only in a part of the 
filament, and seems to be a standing oscillation rather than 
a propagating wave. 
The spatial displacement of the oscillation was also identified 
in 17 GHz microwave images from NoRH. 
Such an oscillation has not been reported during the more than 
ten years of NoRH observations (although we have not done a systematic survey 
for such events, so this does not mean that ours is the only such event), 
and we believe that this is the first observation of spatially resolved 
filament oscillation by microwave imaging. 

In Contrast to previous observations of large-amplitude filament 
oscillations, there were no flares or even microflares associated 
with the excitation of the filament. However, we found that a 
filament barb connected near an emerging flux disappeared 
and seems to be ejected at the onset of the oscillation. 
We believe that reconnection between the emerging flux 
and the barb caused the ejection that blew the filament 
to excite the oscillation. 

The large-amplitude oscillation occurred in the slow-rise phase of the filament, 
and only a part of the filament exhibited the oscillation. 
The oscillation is evidence that the oscillating part of the filament still retained 
a nonlinearly stable equilibrium during the slow rise. 
It supports the model by Chifor \ietal (2006) that 
at first a part of filament loses equilibrium and starts to rise, and the slow rise 
induces LOE of the remaining part, leading to the 
destabilization of the whole filament. 

Filament and prominence seismology is a powerful tool, not only for 
measuring the physical parameters, but also for diagnosing 
the stability and hence the trigger mechanism of filament eruption. 
It is worthwhile to search for more events in which 
filaments exhibit large-amplitude oscillation in their 
pre-eruption phase. 
From a theoretical point of view, there are many analytical and 
numerical models of pre-eruption filaments or flux ropes.
Investigation of the behavior of these model filaments 
as oscillators may provide some physical insight into 
the mechanisms of filament eruptions.

\acknowledgements
We thank M. Shimojo for fruitful discussions. 
This work was supported in part by a Grant-in-Aid for 
Creative Scientific Research of the Ministry of Education, 
Culture, Sports, Science and Technology
(MEXT) of Japan ``The Basic Study of Space Weather Prediction"
(17GS0208, K. Shibata, Principal Investigator).
HI is supported by a Research Fellowship from the Japan 
Society for the Promotion of Science for Young Scientists. 
DT acknowledges the support from PPARC. 
RJ acknowledges The Nuffield Foundation (UK) for NUF-NAL 04 award.

\begin{figure}    
\begin{center}
\psfig{file=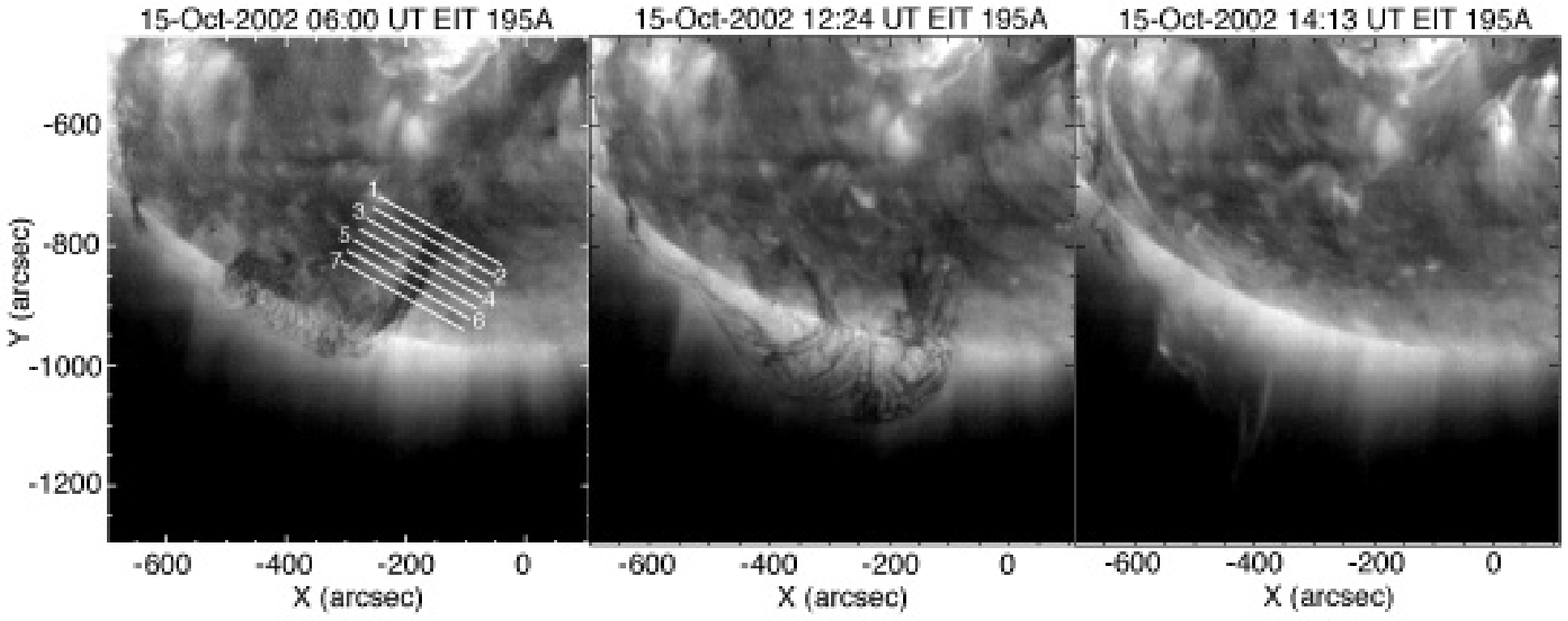,width=12.cm}
\caption{EIT images of the filament during the oscillation 
(left), at the onset of eruption (middle) and after the eruption (right). 
The $x$ and $y$ axes are the distances from the disk center. North is up, 
and East is to the left. 
The solid lines in the left panel with numbers indicate the locations 
of the slits for the time slices shown in Figure 2. 
}
\label{mdi-evol}
\end{center}  
\end{figure}

\begin{figure}    
\begin{center}
\psfig{file=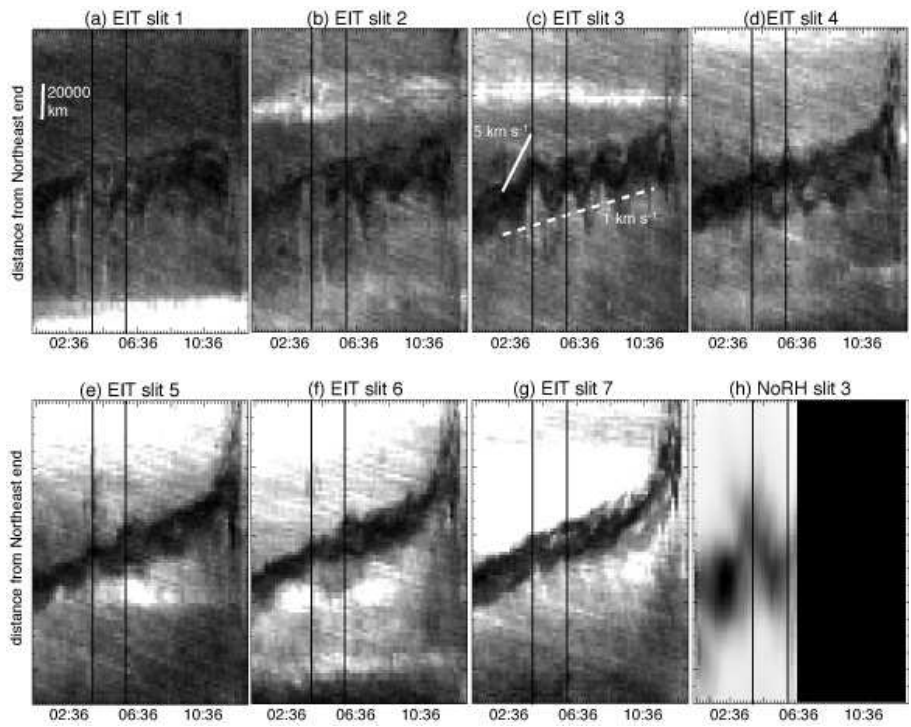,width=12.cm}
\caption{Time slices of EIT (a-g) and NoRH (h) images. 
The positions of the slits are indicated in Figure 1.
The solid and dashed lines in panel (c) indicate 
velocities of 5 and 1 km s $^{-1}$, respectively. 
The black vertical lines indicate 04:00 UT and 06:00 UT, 
as the reference of the first and second peaks of 
the oscillation determined by visual inspection of 
panel (c). 
}
\label{overview}
\end{center}  
\end{figure}
 
\begin{figure}    
\begin{center}
\psfig{file=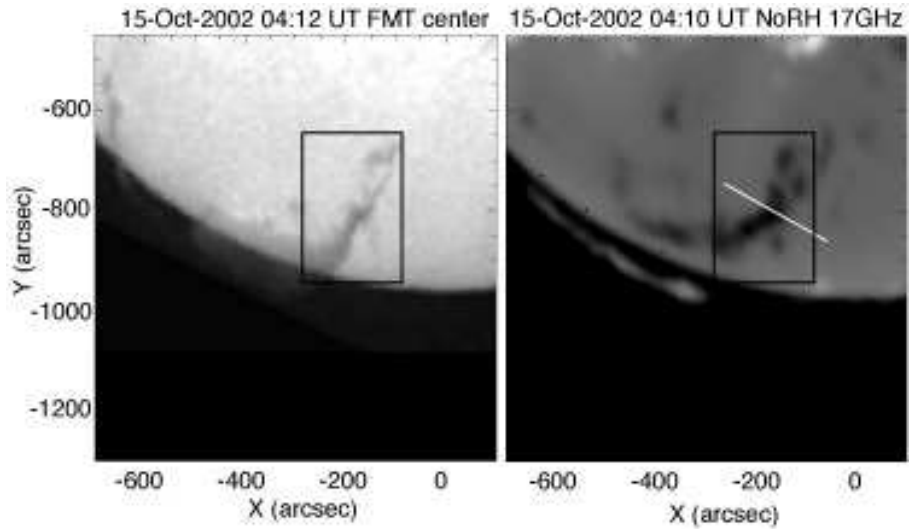,width=12.cm}
\caption{H$\alpha$ center image of the filament 
taken by FMT (left) and 17 GHz microwave image 
taken by NoRH. The field of view is the same as in Figure 1. 
The rectangle shows the field of view of Figure 4. The 
white line on the microwave image indicates the slit position 
of the time slice shown in panel (h) of Figure 2, which 
is the same as that of slit 3. 
}

\label{rfmaps}
\end{center}  
\end{figure}

\begin{figure}    
\begin{center}
\psfig{file=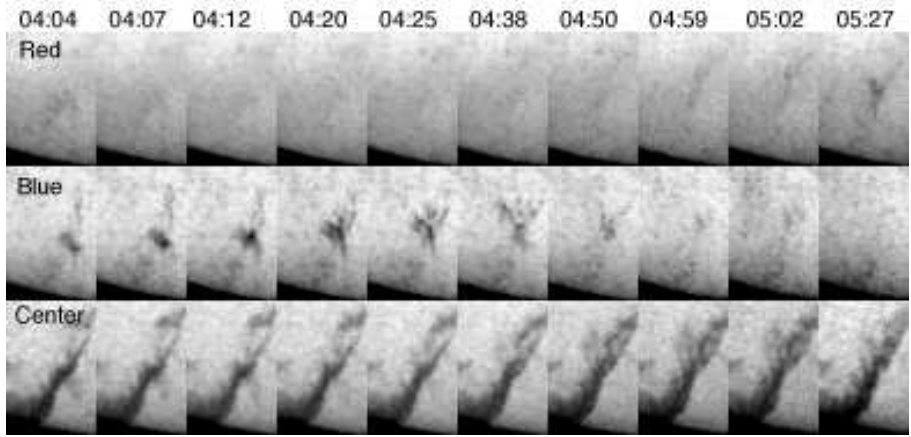,width=12.cm}
\caption{Time series of H$\alpha$ images at the 
line center (bottom row), blue wing at -0.8 \AA \, (middle row), 
and red wing at +0.8 \AA \, (top row). 
The field of view is shown in Figure 3. 
}
\label{fmtimages}
\end{center}  
\end{figure}

\begin{figure}    
\begin{center}
\psfig{file=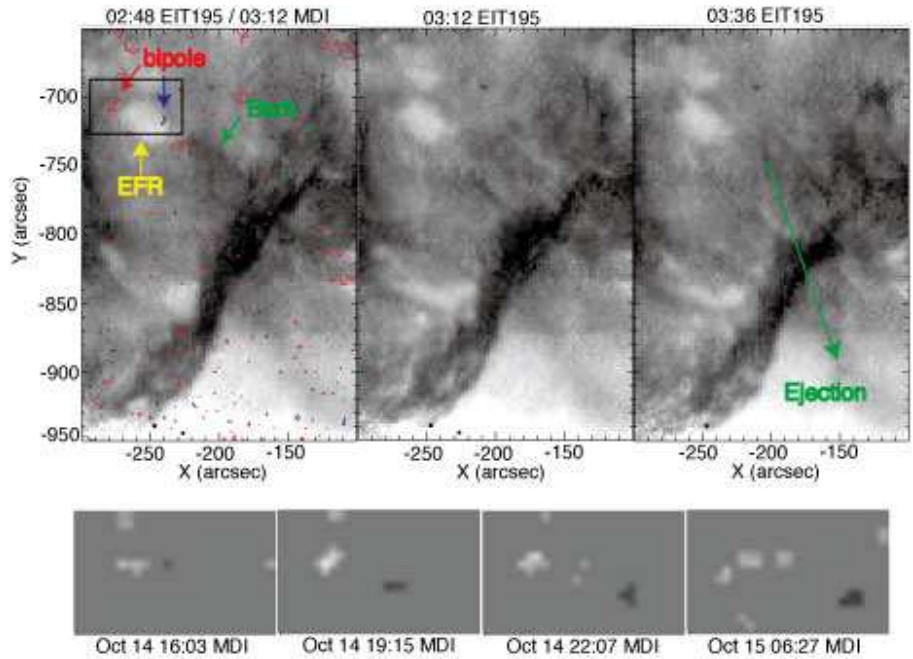,width=12.cm}
\caption{Upper panels: EIT images just before the onset of the oscillation. 
Overlaid on the left panel is the MDI magnetogram taken at 
03:12 UT. The red and blue contours are +25 and -25 gauss, 
respectively. 
Lower panels: Temporal evolution of the MDI magnetograms. The field of view is 
the same as the box in the upper left panel. The pixels in which $|B| < 25$ G 
are set to zero to aid visualization. 
}
\label{efr}
\end{center}  
\end{figure}

\end{article} 
\end{document}